\documentclass[10pt,twocolumn,letterpaper]{article}

\usepackage{cvpr}              

\usepackage{graphicx}
\usepackage{amsmath}
\usepackage{amssymb}
\usepackage{booktabs}

\usepackage{bbm}
\usepackage{bm}
\usepackage{amsthm}
\usepackage[ruled]{algorithm}
\usepackage{algorithmic}
\usepackage{xcolor}
\usepackage{xspace}
\usepackage{setspace}
\usepackage{soul,lipsum}
\usepackage[pagebackref,breaklinks,colorlinks]{hyperref}

\usepackage[capitalize]{cleveref}
\crefname{section}{Sec.}{Secs.}
\Crefname{section}{Section}{Sections}
\Crefname{table}{Table}{Tables}
\crefname{table}{Tab.}{Tabs.}
\def\cvprPaperID{*****} 
\def\confName{CVPR}
\def\confYear{2022}

\newcommand{\Sum}[1]{\raisebox{0.5ex}{\scalebox{0.8}{$\displaystyle \sum_{#1}\;$}}}

\newcommand{\myul}[2][black]{\setulcolor{#1}\ul{#2}\setulcolor{black}}

\begin{document}

\title{Towards Generalisable Audio Representations for Audio-Visual Navigation}

\author{Shunqi Mao \qquad Chaoyi Zhang \qquad Heng Wang \qquad Weidong Cai \\
{School of Computer Science, University of Sydney, Australia}\\
{\tt\small \{smao7434, czha5168, hwan9147\}@uni.sydney.edu.au, tom.cai@sydney.edu.au}
}
\maketitle

\begin{abstract}
    In audio-visual navigation (AVN), an intelligent agent needs to navigate to a constantly sound-making object in complex 3D environments based on its audio and visual perceptions.
    While existing methods attempt to improve the navigation performance with preciously designed path planning or intricate task settings, none has improved the model generalisation on unheard sounds with task settings unchanged.
    We thus propose a contrastive learning-based method to tackle this challenge by regularising the audio encoder, where the sound-agnostic goal-driven latent representations can be learnt from various audio signals of different classes. 
    In addition, we consider two data augmentation strategies to enrich the training sounds.
    We demonstrate that our designs can be easily equipped to existing AVN frameworks to obtain an immediate performance gain 
    (13.4\%$\uparrow$ in SPL on Replica and 12.2\%$\uparrow$ in SPL on MP3D).
    Our project is available at https://AV-GeN.github.io/.
\end{abstract}

\section{Introduction}
\label{sec:intro}
Intelligent robots ought to be capable of perceiving, interpreting, and acting, given the multi-sensory signals from the surrounding environment.
The recent SoundSpaces challenge \cite{soundspace} provides a desired testbed for such ability,
where an embodied agent explores complex 3D environments and finds the positions of sounding objects based on visual and auditory perceptions,
together with a feasible baseline AV-NAV to learn step-by-step actions for navigation with multi-modality inputs.
In \cite{looklisten}, the authors decomposed AVN into predicting the relative position of the sound source and navigating to the position given visual inputs. The AV-WaN \cite{waypoint} further enabled agents to navigate through intermediate waypoints with an occupancy map. 
However, these attempts generalise poorly to unheard sounds. 
Although learning in complex scenarios with multiple non-stationary audio sources could enhance the generalisation \cite{catch}, it requires complicated reformulations of the environment.
More specifically, several works have been proposed to design more complicated tasks to facilitate navigation, such as distractor attacks \cite{adversarial_nav}, short-duration sounds \cite{semantic}, exploration \cite{explore}, and sound separation \cite{move2hear}, while none of them deal with the overfitting problem on training sounds.

To reduce the generalisation errors, we propose a novel Audio Feature Similarity Optimisation (AFSO) method, where the audio features of distinct sounds, which carry identical goal-driven information, will be optimised to be closer in latent space. Moreover, we augment the sounds to prevent learning biased audio encoders. Our method can be conveniently applied to any learning-based AVN paradigms without substantially altering the task definition. 
Evaluated on two large-scale benchmarking datasets Replica \cite{replica19arxiv} and MP3D \cite{Matterport3D}, our method outperforms the existing frameworks by a large margin on the 
\href{https://soundspaces.org/challenge}{\color{blue} \myul[blue] {SoundSpaces}} challenge.

\section{Method}

\subsection{Audio Feature Similarity Optimisation}
\begin{figure*}
  \centering
  \begin{subfigure}[t]{0.43\linewidth}
    \includegraphics[width=1\linewidth]{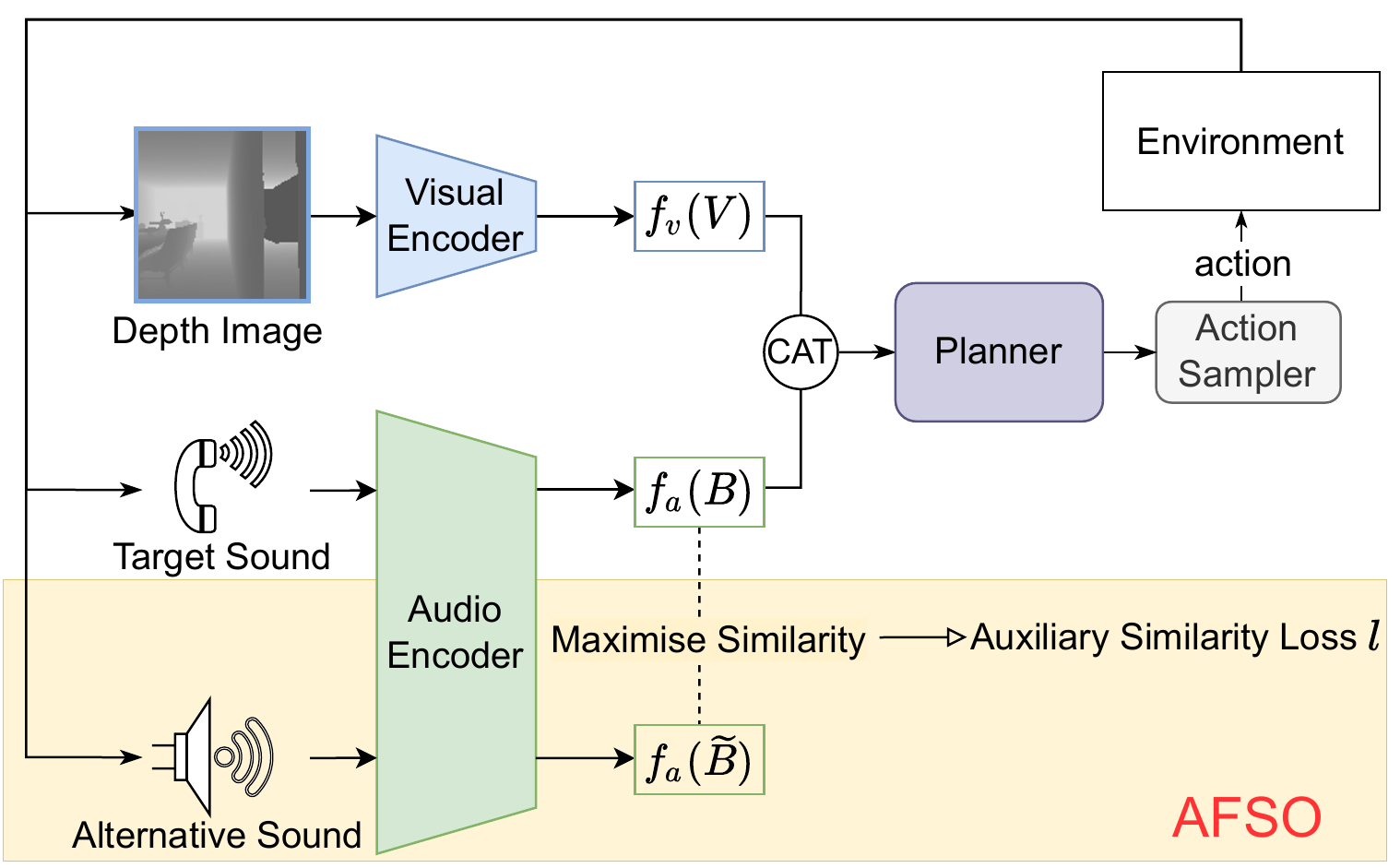}
    \caption{}
    \label{fig::afso_framework}
  \end{subfigure}
  \hfill
  \begin{subfigure}[t]{0.56\linewidth}
    \includegraphics[width=\linewidth]{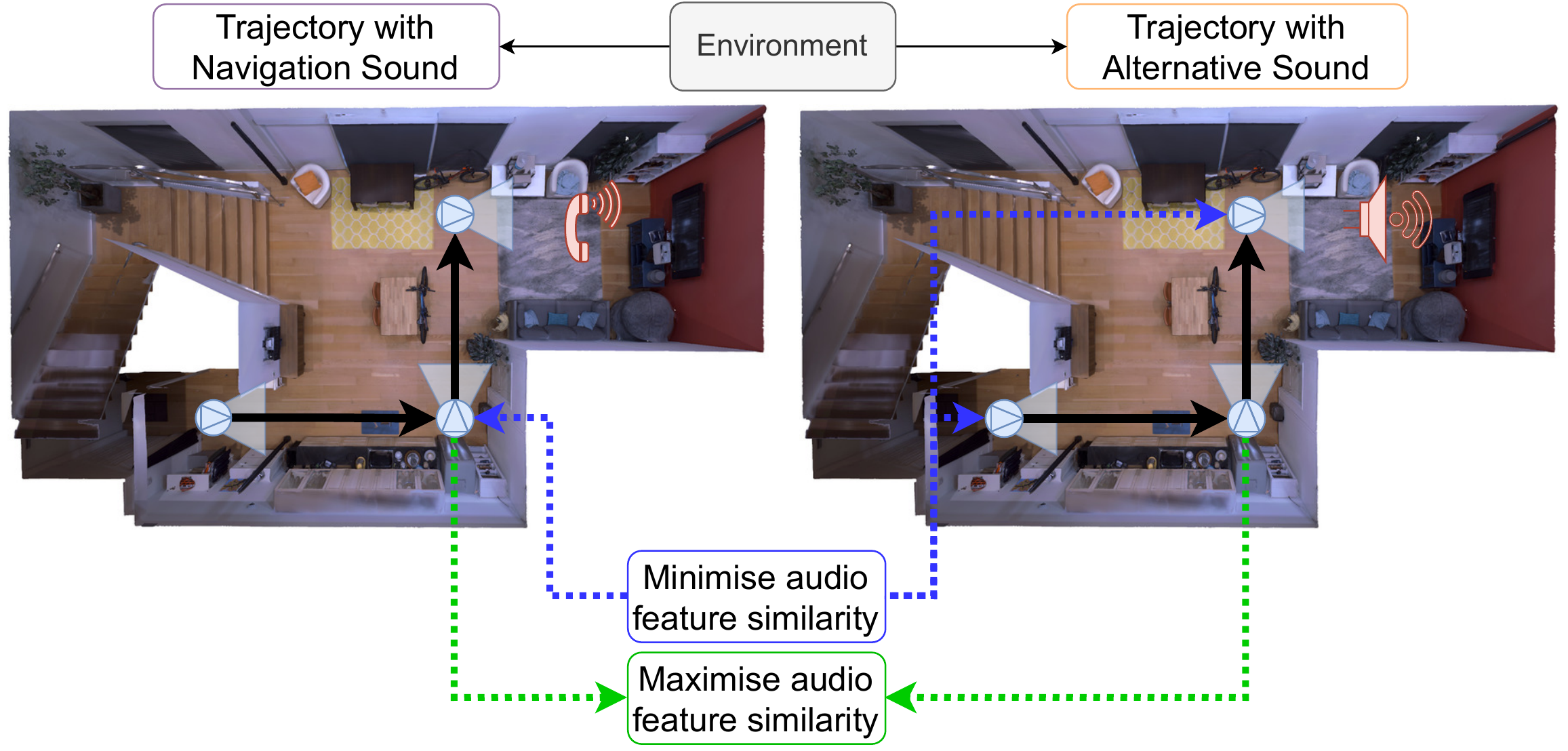}
    \caption{}
    \label{fig::trajectory_view}
  \end{subfigure}
  \caption{Our proposed audio feature similarity optimisation (AFSO) method. (a) Schematic illustration of how our AFSO method is plugged into a generic AVN framework. (b) The graph shows two identical trajectories with telephone and speaker as source sounds. The AFSO method maximises the feature similarity of positive audio-pairs of equivalent source-receiver spatial relationships and minimises the similarity between the negative pairs.
  }
  \label{fig::afso}
\end{figure*}

Modern AVN frameworks encode the acoustic signals on the binaural audio spectrograms. 
While aiming to learn the source-receiver spatial relationships (\ie, the positions of sounding objects w.r.t. the agent), they are easily subject to the distinction of different sound types (\eg, telephone or speaker). 
Inspired by recent contrastive learning strategy \cite{simclr}, we propose the Audio Feature Similarity Optimisation (AFSO) method to alleviate the effects of sound types and focus on learning spatial relationships.
Guiding the audio encoder with auxiliary similarity loss as depicted in \cref{fig::afso_framework}, we explicitly maximise the feature similarity between two audio observations that are sourced from different sounds but at the same relative position to the agent, and minimise the one between two audio observations at different relative sounding positions. In this way, the audio encoder can now focus on goal-driven patterns that are indicative of the target position without overfitting to specific sounds.

Specifically, a pair of binaural audio signals are considered similar only if they are sourced from the identical emitting-receiver positions within the same scene. Such two audio signals will form a positive pair, whereas all other pairs will be reckoned as negative samples. We illustrate an example in \cref{fig::trajectory_view}.
To efficiently form the training pairs, we directly simulate the second element in the pair instead of searching for the matched ones from collected trajectories.
For each binaural audio signals $b_k$ in a training batch, we simulate positive pairing acoustic signal $\widetilde{b}_k$ by convolving the room impulse response at the current step with an alternative type of source sound. The audio data is then transformed to binaural spectrograms as inputs to the audio encoder.
However, such a formulation could potentially introduces false-negative (FN) pairs, where identical or similar audio observations in the trajectories may be treated as negative samples. To reduce the occurrence of FN pairs, we only compute the alternative audio and the similarity loss for a randomly sampled subset of $N$ audio observations from the trajectories. Therefore, we obtain $2N$ data samples per batch including $N$ original sounds and $N$ corresponding simulated audio signals. Following \cite{simclr}, for a positive pair of audio observation $(b_i, b_j)$, we calculate the auxiliary similarity loss using InfoNCE loss \cite{cpc} as $\ell_{i,j} = -\log \frac{\exp(\mathrm{sim}(f_a(b_i), f_a({ b_j)}/\tau)}{\sum_{k=1}^{2N} \mathbbm{1}_{k \neq i}\exp(\mathrm{sim}(f_a( b_i), f_a({ b_k)}/\tau)}~$,
where $f_a$ represents the audio encoder, $\mathrm{sim}$ denotes the cosine similarity function 
$\mathrm{sim}(\bm u,\bm v) = \bm u^\top \bm v / \lVert\bm u\rVert \lVert\bm v\rVert$, and $\tau$ is a temperature parameter.
We apply a weight factor $w$ to the similarity loss $L=\Sum{B} l_{i,j}$ and combine it with standard AVN losses for optimisation.

\subsection{Source Sound Augmentation}
To enrich the training sounds distribution, we also apply two sound augmentation strategies. At each episode, we augment the source sounds with the following techniques: 1) Sound Reverse: We reverse the input audio signals with a probability of $p$; 2) Sound Mix-up: We sample two audios signals (potentially reversed) and mix them as $x_{mix} = \lambda x_1 + (1 - \lambda) x_2$, where $\lambda$ is a scalar sampled from the symmetric Beta distribution $\lambda \sim Beta(\alpha, \alpha)$. 

\section{Experiments}
We evaluate our method by deploying it to two state-of-the-art audio-visual models, AV-NAV \cite{soundspace} and AV-WaN \cite{waypoint}, on two benchmarking datasets, Replica \cite{replica19arxiv} and MP3D \cite{Matterport3D}, following their official dataset split \cite{soundspace}.
We reproduce two baselines with default hyperparameters specified in papers.
For our AFSO and sound augmentation strategies, we set the weight factor $w$ as $0.1$, the temperature $\tau$ as $0.07$, the audio batch $N$ as $256$, the reverse probability $p$ as $0.5$, and the mix-up factor $\alpha$ as $1$.
We report SPL, SR and SNA as evaluation metrics \cite{soundspace, waypoint}.

We evaluate all methods on test splits with \textit{unseen} scenes and \textit{unheard} sounds. As presented in \cref{tab:result},   
our method significantly improves the performance of the baseline models on both datasets with around 12\% growth in SPL, which demonstrates our methods' ability in improving the generalisation of the audio encoder and preventing overfitting. 

\begin{table}[!htp]\centering
\scriptsize
\begin{tabular}{l|ccc|ccc}\toprule
&\multicolumn{3}{c|}{\textit{Replica}} &\multicolumn{3}{c}{\textit{MP3D*} } \\ \cmidrule(lr){2-7}  
&SPL$\uparrow$ &SR$\uparrow$ &SNA$\uparrow$ &SPL$\uparrow$ &SR$\uparrow$ &SNA$\uparrow$ \\\midrule
AV-NAV &38.2 &45.2 &21.5 &26.3 &43.6 &11.8 \\
AV-NAV + Ours &\textbf{51.4} &\textbf{64.4} &\textbf{30.4} &\textbf{37.1} &\textbf{55.8} &\textbf{19.1} \\\midrule
AV-WaN &35.7 &48.4 &28.5 &36.2 &57.4 &27.4 \\
AV-WaN + Ours &\textbf{49.1} &\textbf{69.8} &\textbf{38.6} &\textbf{48.4} &\textbf{73.9} &\textbf{36.9} \\
\bottomrule
\end{tabular}
\caption{Quantitative comparisons with AVN SOTAs in test splits on Replica and MP3D (* denote the SoundSpaces challenge \cite{soundspace}).
}\label{tab:result}
\end{table}

\section{Conclusion}
We introduce a contrastive learning-based approach AFSO to regularise the audio encoder to distill goal-driven representations for AVN. Our method can be easily deployed to existing frameworks, and substantially enlarges their generalisation ability on unfamiliar sound sources.

{\small
\bibliographystyle{ieee_fullname}
\bibliography{main}
}

\end{document}